\documentclass[aps,prd,twocolumn,superscriptaddress,nofootinbib,showpacs,10pt]{revtex4-1}
\usepackage[utf8]{inputenc}
\usepackage[english]{babel}
\usepackage{amsmath}
\usepackage{amsfonts}
\usepackage{amssymb}
\usepackage{graphics}
\usepackage{graphicx}
\usepackage[left=2cm,right=2cm,top=2cm,bottom=2cm]{geometry}

\usepackage[usenames,dvipsnames,svgnames]{xcolor}

\begin{document}
\title{Amplified transduction of Planck-scale effects using quantum optics}

\author{Pasquale Bosso} \email{pasquale.bosso@uleth.ca}
\author{Saurya Das} \email{saurya.das@uleth.ca}
\affiliation{Theoretical Physics Group and Quantum Alberta, University of Lethbridge,\protect\\ 4401 University Drive, Lethbridge, Alberta, Canada, T1K 3M4}
\author{Igor Pikovski} \email{igor.pikovski@cfa.harvard.edu}
\affiliation{ITAMP, Harvard-Smithsonian Center for Astrophysics, Cambridge, MA 02138, USA}
\affiliation{Department of Physics, Harvard University, Cambridge, MA 02138, USA}
\author{Michael R. Vanner} \email{michael.vanner@physics.ox.ac.uk}
\affiliation{Clarendon Laboratory, Department of Physics, University of Oxford, OX1 3PU, United Kingdom}

\date{\today}

\begin{abstract}
The unification of quantum mechanics and gravity remains as one of the primary challenges of present-day physics. Quantum-gravity-inspired phenomenological models offer a window to explore potential aspects of quantum gravity including qualitatively new behaviour that can be experimentally tested. One such phenomenological model is the generalized uncertainty principle (GUP), which predicts a modified Heisenberg uncertainty relation and a deformed canonical commutator. It was recently shown that optomechanical systems offer significant promise to put stringent experimental bounds on such models. In this paper, we introduce a scheme to increase the sensitivity of these experiments with an extended sequence of pulsed optomechanical interactions. We also analyze the effects of optical phase noise and optical loss and present a strategy to mitigate such deleterious effects.
\end{abstract}


\maketitle

\section{Introduction}
At present there is no successful theory of quantum gravity. One route to finding such a theory is through the study of phenomenological models, which can motivate conceptual and experimental studies of the interplay between quantum mechanics and general relativity~\cite{ACamelia1, Hossenfelder1, Plato2016}. Experimental bounds on a variety of such models have been obtained, both from the observed validity of quantum mechanics in a variety of systems and from studies focussed on experimentally testing for such phenomena~\cite{expQGR, dv, Marin2013_1, Marin2014, Bawaj2015, Bekenstein2012_1}. Although, thus far, there has been no experimental evidence of any modifications to quantum mechanics or general relativity. An important task to guide further theoretical development in this area is to obtain improved experimental bounds that have the potential to rule out some models and/or restrict where free parameters can lie.

A specific class of phenomenological models assumes that the Heisenberg uncertainty principle undergoes modifications near the Planck energy or the Planck length. This so-called generalized uncertainty principle (GUP) is motivated by various quantum gravity models \cite{Garay}, including string theory \cite{guppapers1}, micro black holes \cite{Scardigli}, doubly special relativity \cite{cg}, and Gedankenexperiments at short distances \cite{ACamelia1, Hossenfelder1, Garay}. Because of this broad applicability, a quantum theory based on the GUP has been the subject of a wide range of studies \cite{guppapers1,maggiore,kmm,guppapers2,Ali2011_1}. One of the main implications is the lack of position eigenstates and the appearance of a minimal-length scale \cite{kmm}. Effects of the GUP on various quantum systems and phenomena have been studied, such as a particle in a box, the simple harmonic oscillator, the Lamb shift, scanning tunnelling microscopy \cite{dv,Ali2011_1}, as well as its effects in cosmology \cite{barun}. It was also shown that the GUP implies discreteness of measured lengths, areas and volumes at the fundamental level \cite{ADV,Das2010_1,Deb2016_1}. In the last few years, several schemes to test the GUP have been proposed. These include optomechanical interactions~\cite{Pikovski2012_1}, gravitational bar detectors~\cite{Marin2013_1}, nanodiamond interferometry~\cite{Albrecht2014}, and direct measurements on a harmonic oscillator~\cite{Rossi2016}.

Several different forms of the GUP have been proposed in the literature \cite{maggiore,  guppapers2,kmm,cg,dv,Ali2011_1}. A general form of the GUP, expressed in terms of an inequality and in terms of a deformed commutator, is given, to $\mathcal{O}(\gamma^2)$, by
\begin{align}
	\begin{split}
	 \Delta x \Delta p  & \geq  \frac{\hbar}{2} \left[ 1 - 2 \gamma \langle p \rangle + 4\gamma^2 \langle p^2 \rangle \right], \\
	 \geq  \frac{\hbar}{2}  & \left[ 1  +  \left(  \frac{\gamma}{\sqrt{\langle p^2 \rangle}} + 4\gamma^2 \right)  \Delta p^2 +  4\gamma^2 {\langle p \rangle}^2  -  2\gamma \sqrt{\langle p^2 \rangle} \right] , \label{dxdp1}
	\end{split}\\
	[x_i, p_j] = & i \hbar \left[  \delta_{ij} -  \gamma \left( p \delta_{ij} + \frac{p_i p_j}{p} \right) + \gamma^2 \left( p^2 \delta_{ij}  + 3 p_{i} p_{j} \right) \right], \label{comm1}
\end{align}
where the indices $i$ and $j$ capture the three spatial components and $p^2=\sum_i^3 p_i^2$. Here, $\gamma = {\gamma_0}/{M_\textsc{PL}c} = {\gamma_0 \ell_\textsc{PL}}/{\hbar}$ is the GUP parameter in terms of inverse Planck momentum, and $\gamma_0$ is the dimensionless deformation strength. (The Planck units are the Planck mass $M_\textsc{PL}\approx 22$~$\mu$g, the Planck length $\ell_\textsc{PL}\approx 10^{-35}$~m, and the Planck energy $M_\textsc{PL} c^2 \approx 10^{19}$~GeV.) For most models it is anticipated that these deformations arise at the Planck scale, i.e. $\gamma_0 \simeq 1$. The quadratic GUP proposed in Refs.~\cite{maggiore,kmm} is a special case of the above, when the term linear in $\langle p \rangle$ is set to zero.
In Ref. \cite{Maggiore1993_2} a type of GUP was proposed based on algebraic structure, from which the quadratic terms in momentum and higher orders follows the commutator
\begin{eqnarray}
[x_i, p_j] = i\hbar \delta_{ij} \sqrt{1 + \gamma^2 \left[ p^2 + (mc)^2 \right] } \label{comm2},
\end{eqnarray}
where $m$ denotes the mass of the quantum object. In the following, we will be working in one spatial dimension (with coordinate $x$ and momentum $p$), applying the various GUPs to a quantum mechanical oscillator.

In Ref.~\cite{Pikovski2012_1} it was shown that optomechanics can be used to tightly constrain the free parameters in GUP models, under the assumption that the centre-of-mass motion is affected. Here we present an extension to this scheme and show that multiple pulsed interactions can be used to improve the sensitivity of the experiment. In particular, we analyze optical phase noise, optical loss, and mechanical decoherence, identifying the parameter regime where this extension is advantageous. We first present the scheme and main ideas of Ref. \cite{Pikovski2012_1}. Then in Section \ref{ncycles} we describe our scheme to amplify the signal using multiple cycles. In Section \ref{sec:deleterious_effects} we analyze optical phase noise, optical loss and mechanical decoherence. Exploiting the multiple pulse sequence proposed here, we also present a strategy in this section to reduce the unwanted effects of optical loss.

\section{Pulsed Optomechanical scheme to probe GUP}

The basic idea presented in Ref.~\cite{Pikovski2012_1} is to utilise a sequence of four radiation-pressure interactions with an optical pulse and a mechanical harmonic oscillator (mass: $m$; angular frequency: $\omega_\textsc{m}$), which forms one end-mirror of an optical cavity~\cite{Law1995_1}. The scheme allows one to precisely infer the value of the mechanical canonical commutator, including possible quantum-gravity-induced deformations, by measuring how the phase of the light field is changed by the interaction. Introducing the dimensionless mechanical position and momentum quadrature operators
\begin{subequations}
	\begin{align}
		X_\textsc{m} &= x \sqrt{\frac{m \omega_\textsc{m}}{\hbar}};\\
		P_\textsc{m} &= \frac{p}{\sqrt{\hbar m \omega_\textsc{m}}},
	\end{align}
\end{subequations}
one can show that in a pulsed regime~\cite{Vanner2011} the interaction can be described in terms of the unitary operator
\begin{equation}
	U_{\textsc{lm}} = e^{i\lambda {n}_\textsc{l} {X}_\textsc{m}},
\end{equation}
where $\lambda$ represents the interaction strength, and $n_\textsc{l}$ is the photon number operator. Note that $[ X_\textsc{m},  P_\textsc{m}]=[ x, p]/\hbar$. In what follows, the subscript M is used to identify mechanical quantities, and the subscript L is used for optical degrees-of-freedom.

After the first interaction, the same pulse re-enters the cavity after a quarter of a mechanical period and this is repeated three more times to complete a full mechanical cycle. The total interaction is described by the operator
\begin{equation}
{\xi} = e^{i\lambda {n}_\textsc{l} {P}_\textsc{m}} e^{-i\lambda {n}_\textsc{l} {X}_\textsc{m}} e^{-i \lambda {n}_\textsc{l} {P}_\textsc{m}} e^{i \lambda {n}_\textsc{l} {X}_\textsc{m}}. \label{eqn:four-displacement_operator}
\end{equation}
The operator ${\xi}$ contains information on the commutation relation between ${X}_\textsc{m}$ and ${P}_\textsc{m}$ and imprints this information onto the optical field. To quantify this change to the optical field we examine the mean of the field operator, i.e.
\begin{equation}
	\langle {a} \rangle = \langle \bar \alpha | {\xi}^\dagger {a} {\xi} | \bar \alpha \rangle
=\langle {a} \rangle_\textsc{qm} e^{-i\Theta},
\end{equation}
where ${a}$ is the annihilation operator for the optical field and
$| \bar \alpha\rangle$ is the input optical coherent state with mean photon number $N_\textrm{p}$. The mean of the field operator $a$ using standard quantum mechanics is $\langle {a} \rangle_\textsc{qm} =\bar \alpha~e^{-i\lambda^2 - N_\textrm{p}(1 - e^{-i2\lambda^2})}$, and the (complex) quantity $\Theta$ describes the Planck-scale-physics-dependent additional contribution, which to lowest order is just an optical phase shift. The real and imaginary parts of ${a}$ are the observable amplitude and phase quadratures of the light, respectively. It is important to note that the above expression does not depend on the mechanical state of motion. This is because Eq. \eqref{eqn:four-displacement_operator} reduces to ${\xi} = e^{-i\lambda^2 {n}_\textsc{l}^2}$ in the quantum mechanical case, and also to leading order in Planck-scale deformations, the mechanical operators drop out of the expression. As optomechanical systems operate in the regime $\lambda \ll 1$ to date, the leading order change to $\langle {a} \rangle$ is a change in the optical phase. (For couplings $\lambda \ge 1$, the mean of the quadratures is suppressed by $e^{-N_p(1-\textsc{cos}(2\lambda^2))}$, since higher order moments are also affected).

For the purely quadratic GUP, and for the models in (\ref{comm1}), and (\ref{comm2}), we have
\begin{subequations}
	\begin{align}
		\Theta &\simeq \frac{4}{3} \bar{\gamma}^2 N_\textrm{p}^3 \lambda^4 e^{-i6\lambda^2}; \\ 
		\Theta &\simeq \frac{3}{2} \bar{\gamma}^2 N_\textrm{p}^2 \lambda^3 e^{-i4\lambda^2}; \\ 
		\Theta &\simeq \bar{\gamma}^2 m^2 c^2 N_\textrm{p} \lambda^2 e^{-i2\lambda^2},
	\end{align}
\end{subequations}
respectively, where $\bar{\gamma} = \gamma \sqrt{\hbar \omega_\mathrm{M} m}$. Note the different nonlinear scalings with $N_\textrm{p}$ and $\lambda$ between these three predicted corrections.

An analysis has also recently been published~\cite{Armata2016} that determined what fraction of $\langle {a}\rangle_\textsc{qm}$ can be described classically, thus there is a hierarchy between classical, quantum, and quantum-gravitational predictions.

\section{Extending the sequence of optomechanical interactions}
\label{ncycles}

The above approach opens a new avenue to measure Planck-scale effects in a low-energy regime on a table-top and it is highly desirable to enhance the sensitivity of the scheme to more deeply explore the GUP parameter regime. Also, the coupling strengths currently available in solid-state implementations of optomechanics are small, i.e. $\lambda \ll 1$, which limit the signal strength. In this section, we show that such amplifications are possible by repeating the four-pulse-sequence over $N$-cycles. As can be seen from  Eq.~(\ref{eqn:four-displacement_operator}), after $N$-cycles, the four-pulse operator would simply be $\xi^N$, when neglecting deleterious effects such as changes in the pulse shape, optical losses, or decoherence. In the following sub-sections, we compute this operator for standard quantum mechanics, the linear and quadratic GUP, only quadratic, and the higher-order GUPs. In all of these cases we assume that the photon number per pulse is large.

\subsection{Standard quantum mechanical prediction}
\label{ssec:HUP_no_mods}

Since it follows from the unmodified $[X_\textsc{m},P_\textsc{m}]$ commutator that ${\xi} = e^{-i\lambda^2 {n}_\textsc{l}^2}$,
after $N$-cycles, one gets
\begin{equation}
	{\xi}^N = e^{-iN\lambda^2 {n}_\textsc{l}^2}. \label{eqn:canonical_xi}
\end{equation}
Similarly, using the Hausdorff formula, one gets
\begin{equation}
	\left({\xi}^N\right)^\dagger {a} {\xi}^N = e^{-iN\lambda^2(2 {n}_\textsc{l} + 1)} {a}~,
\end{equation}
and finally, using the following well known properties of coherent states
\begin{eqnarray}
	{a} |\alpha\rangle &=& \alpha | \alpha \rangle;\\
	e^{i\varphi {n}_\textsc{l}}|\alpha\rangle &=& |\alpha e^{i\varphi}\rangle;\\
	\langle\alpha|\beta\rangle &=& e^{-\frac{|\alpha|^2 + |\beta|^2}{2}} e^{\alpha^*\beta} = e^{-\frac{|\alpha-\beta|^2}{2}}e^{i\Im(\alpha^*\beta)};\\
	\langle {n}_\textsc{l}\rangle &=& \langle\alpha|{n}_\textsc{l}|\alpha\rangle = |\alpha|^2,
\end{eqnarray}
one obtains for the mean optical field
\begin{equation}
	\langle {a}\rangle_{\textsc{qm},N} = \bar\alpha e^{-iN\lambda^2 - N_\textrm{p}\left(1-e^{-i2N\lambda^2}\right)}. \label{eqn:canonical_mean_optical_field}
\end{equation}
Therefore, a small coupling strength $\lambda$ can be compensated by using a large number of loops $N$ to observe the non-classical component of the effect to the light field~\cite{Armata2016}.

\subsection{Linear and quadratic GUP}

For the GUP defined by the commutator (\ref{comm1}), including linear and quadratic terms in momentum, one finds
\begin{align}
	{\xi} =& e^{-i\lambda^2 {n}_\textsc{l}^2} e^{i\bar{\gamma}\left(\lambda^2 {n}_\textsc{l}^2 {P}_\textsc{m} + \frac{1}{2}\lambda^3 {n}_\textsc{l}^3\right)}; \\
	{\xi}^N =& e^{-iN\lambda^2 {n}_\textsc{l}^2} e^{iN\bar{\gamma}\left(\lambda^2 {n}_\textsc{l}^2 {P}_\textsc{m} + \frac{1}{2}\lambda^3 {n}_\textsc{l}^3\right)}.
\end{align}
Carrying on the analysis as in the previous sub-section, one now obtains
\begin{align}
	\left({\xi}^N\right)^\dagger {a} {\xi}^N = & e^{-iN\lambda^2(2{n}_\textsc{l} + 1)}e^{-i \frac{1}{2} N \bar{\gamma} \lambda^3 \left(3{n}_\textsc{l}^2 + 3{n}_\textsc{l} + 1\right)}{a} \, ; \\
	\langle {a} \rangle_{N} \simeq & \langle {a} \rangle_{\textsc{qm},N} e^{-i \Theta(N)}; \\
	\Theta (N) \simeq & \frac{3}{2}N \bar{\gamma} N_\textrm{p}^2 \lambda^3 e^{-i4N\lambda^2} \label{eqn:theta_gamma}~.
\end{align}
Thus, we can see that the desired deformation contribution $\Theta$ is enhanced by a factor of $N$, which contributes only to the phase of the light as long as $N \lambda^2 \lesssim 1$. For sufficiently large $N$, the effect is no longer just a phase and affects higher moments of the optical field. However, comparison to Eq. \eqref{eqn:canonical_mean_optical_field} shows that in this limit the contribution from the deformation scales differently than the regular contribution and can in principle be isolated. The number of loops can be chosen such that the GUP-contribution does not vanish. In practice, however, the number of loops will be limited by deleterious effects, as discussed in section \ref{sec:noise} below.

\subsection{Quadratic GUP}

We now consider the original GUP, which consists of just the quadratic term in momentum in the RHS of (\ref{comm1}).
Following the steps of Ref.~\cite{Pikovski2012_1} for the operator ${\xi}$, we now find for the displacement operator, after $N$ mechanical periods, to leading order
\begin{equation}
	{\xi}^N = e^{-iN\lambda^2 {n}_\textsc{l}^2} e^{-iN \bar{\gamma}^2\left(\lambda^2 {n}_\textsc{l}^2 {P}_\textsc{m}^2 + \lambda^3 {n}_\textsc{l}^3 {P}_\textsc{m} + \frac{1}{3} \lambda^4 {n}_\textsc{l}^4\right)},
\end{equation}
and the mean optical field
\begin{equation}
	\langle {a} \rangle _{N} \simeq \langle {a} \rangle_{\textsc{qm},N} e^{-i\Theta (N)},
\end{equation}
with
\begin{equation}
	\Theta (N) \simeq \frac{4}{3} N \bar{\gamma}^2 N_\textrm{p}^3 \lambda^4 e^{-i6N\lambda^2}. \label{eqn:theta_beta}
\end{equation}

\subsection{Higher order GUP}

Finally, for the GUP associated with Eq.(\ref{comm2}), following the steps of Ref.~\cite{Pikovski2012_1} one has for $N$-cycles (and for $ mc\gg p$):
\begin{align}
	{\xi}^N  = & e^{-iN\lambda^2 {n}_\textsc{l}^2(1+(\bar{\gamma} m c)^2/2)}; \\
	\left({\xi}^N \right)^\dagger {a} {\xi}^N  = & e^{-iN\lambda^2(2 {n}_\textsc{l} + 1)(1+(\bar{\gamma} m c)^2/2)} {a}; \\
	\langle {a} \rangle _{N} = & \bar\alpha e^{-iN\lambda^2(1+(\bar{\gamma} m c)^2/2) - N_\textrm{p}\left[1-e^{-i2N\lambda^2(1+(\bar{\gamma} m c)^2/2)}\right]} \\
& \simeq \langle {a} \rangle_{\textsc{qm},N} e^{-i \Theta(N)}; \\
	\Theta (N) \simeq & N (\bar{\gamma} m c)^2 N_\textrm{p} \lambda^2 e^{-i2N\lambda^2}.
\end{align}
Note that this is also proportional to the inverse square of the Planck mass.

\section{Experimental concerns and the signal-to-noise ratio}\label{sec:noise}

\label{sec:deleterious_effects}

In all experiments there are many sources of noise and imperfections, both of fundamental and technical origins, that can limit performance. We have identified several effects that can reduce the signal-to-noise ratio in our scheme to detect these potential Planck-scale deformations. These noise sources include: optical phase noise (intrinsic quantum noise and additional classical noise), optical loss, mechanical anharmonicity, mechanical decoherence, and classical optical intensity noise. The signal $\Theta$ then has to be compared to the sensitivity of measuring the optical phase shift $\phi$. If we call $\delta \phi$ the uncertainty of measuring $\phi$, then $\Theta  > \delta \phi$ is desirable.

In the following three subsections we discuss optical phase noise, optical loss, and mechanical decoherence, respectively, as we feel that these will be important noise contributions. 
We would like to briefly note here that mechanical anharmonicity, i.e. using a mechanical oscillator with a small Duffing nonlinearity, will also yield nonlinear phase shifts that could be confused with the signals of interest here, see also Ref.~\cite{Latmiral2016}. This effect can be minimized by the choice of mechanical resonator material and by carefully measuring the scaling of the signal with optical intensity.
We would also like to briefly describe the effect of optical intensity noise here and how this can degrade the signal-to-noise ratio. While an optical phase shift is measured at the end of our protocol, which is the conjugate degree of freedom, optical intensity noise has an indirect effect. If the optical intensity fluctuates from one experimental run to the next, the mechanical oscillator will undergo closed loops in phase space of differing sizes and hence the optical field will pick up differing phase shifts. This unwanted effect can be minimized by pre-filtering the input light with an optical cavity as well as employing active intensity stabilisation techniques (`noise eater') to ensure quantum noise limited optical input. In this case the mean of the output optical phase shift is unaffected by such noise, but the phase uncertainty will increase in the presence of optical intensity noise in proportion to $N_\textrm{p}$.
It is beyond the scope of the current work to give a detailed analysis of all of these effects, however, we feel that this section is important to understand how such a multiple-pulse scheme can be implemented as well as its practical limits.

\subsection{Optical phase noise}

In an ideal experiment with a pure coherent state of light with no loss or classical phase noise, the phase uncertainty is given by $\delta \phi = 1/(2\sqrt{N_\textrm{p} N_\textrm{r}})$. Here, $N_\textrm{p}$ is the mean photon number and $N_\textrm{r}$ is the number of independent runs of the experiment. The contribution from $N_\textrm{p}$ is the intrinsic quantum noise of the optical coherent state and the overall phase uncertainty can also be reduced by averaging over $N_\textrm{r}$ experimental runs. See, e.g. Ref.~\cite{BachorBook} for a discussion of optical quantum noise. It is important to compare this uncertainty to the scaling of the signal, which we have shown to increase in proportion to the number of loops, i.e. $\Theta \propto N$. Thus, the signal to noise ratio (SNR), in ideal conditions, becomes
\begin{equation}
\frac{\Theta}{\delta \phi} \propto N \sqrt{N_\textrm{p} N_\textrm{r}} .
\end{equation}
In this case, it is therefore advantageous to use more cycles $N$ over more individual runs $N_\textrm{r}$ of a single-cycle experiment. This is because $N$ directly amplifies the signal, in contrast to the averaging achievable over $N_\textrm{r}$ runs of the experiment.

Using $N$ cycles is advantageous if no deleterious effects are present. However, using several cycles can also increase the noise in the setup. Thus, for some noise sources, the advantage may be negated, leading to an optimal $N$. We anticipate that the primary noise source will be classical phase noise, which will reduce our ability to estimate the optical phase shift. We then have $\delta \phi \sqrt{N_\textrm{r}} = 1/(2\sqrt{N_\textrm{p}}) + \delta \phi_\textrm{c}$, where $\delta \phi_\textrm{c}$ is the classical phase noise. In any practical application, the classical noise $\delta \phi_\textrm{c}$ will increase with photon number and propagation time. At some sufficiently large $N$ or $N_\textrm{p}$, the noise is dominated by $\delta \phi_\textrm{c}$ and the signal-to-noise ratio will reach a maximum. Such classical noise is not fundamental, however, and can be avoided in principle, but it requires significant effort at large photon numbers. Classical optical phase noise for pulsed optomechanical interactions has been discussed in Ref.~\cite{VannerAnnalen} and for a more general discussion see Ref.~\cite{BachorBook}.

\begin{figure}[!ht]
\centering
\includegraphics[width=0.99\columnwidth]{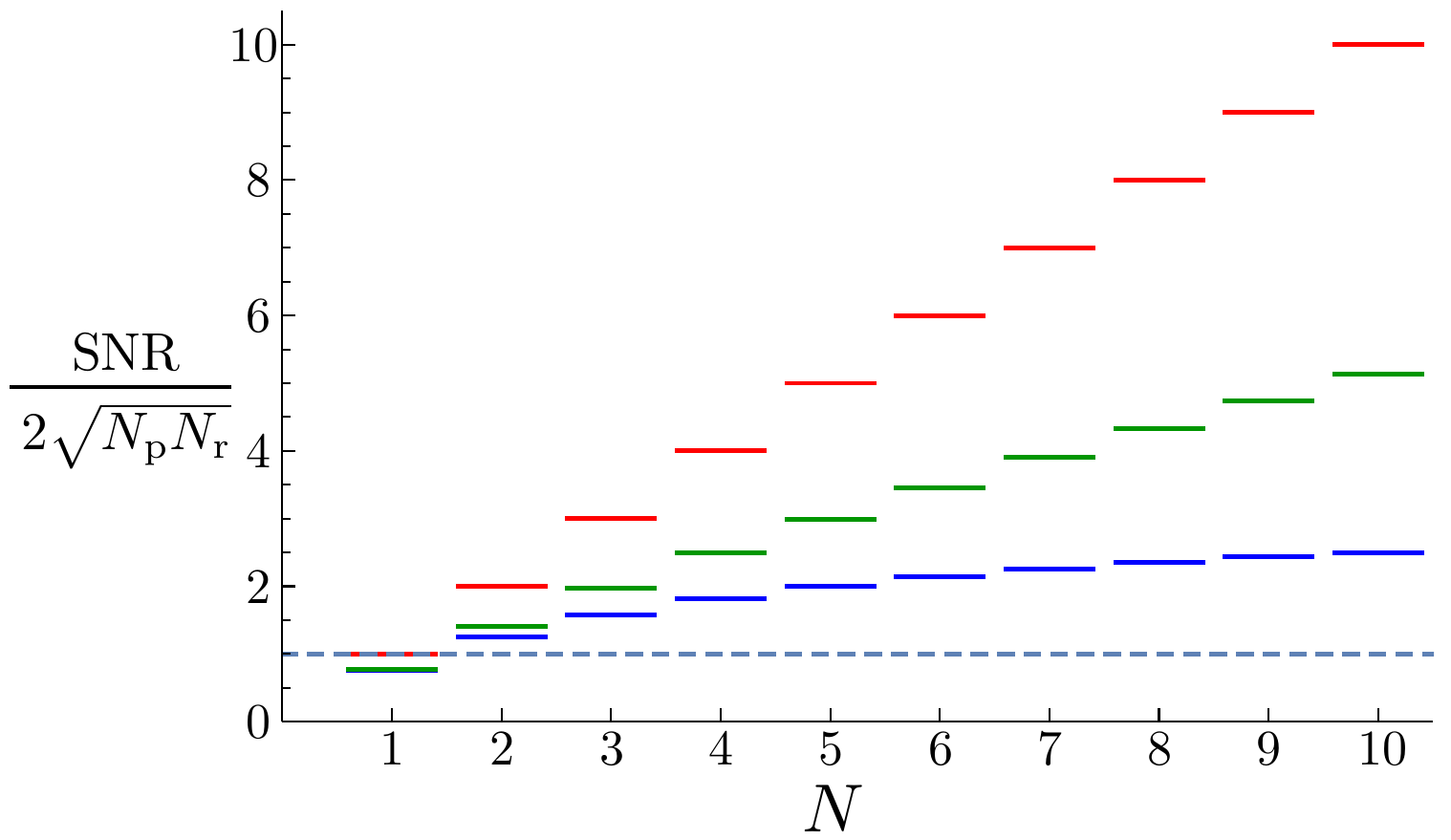}
\caption{\small Plot of signal-to-noise ratio (SNR), Eq.~(\ref{Eq:SNR}), as a function of the cycle number $N$. The upper trace is for the ideal case with no classical noise ($\zeta_\textrm{c} = \zeta_\textrm{u} = 0$, shown in red). The two other curves show the effect of uncorrelated noise ($\zeta_\textrm{c} = 0, \zeta_\textrm{u} = 0.3$, shown in green), and correlated noise ($\zeta_\textrm{c} = 0.3, \zeta_\textrm{u} = 0$, shown in blue). A dashed line at unity is given to aid comparison.}
\label{Fig-SNR}
\end{figure}

Between each interaction, especially within the optical delay loops, and between each run of the experiment there will be fluctuations, e.g. thermal or acoustic, that give a random phase shift to the optical field. We anticipate high-frequency phase noise, which is uncorrelated between loops, and lower frequency noise, which is correlated between loops to both be present but in differing amounts depending on the experimental implementation. Random, uncorrelated noise, will cause the classical noise to scale as $\delta \phi_\textrm{c} \propto \sqrt{N}$ as the phase undergoes a random walk, whereas the scaling of correlated noise may approach $\delta \phi_\textrm{c} \propto N$ as the phase adds between each loop. The signal-to-noise ratio then becomes
\begin{equation}
\frac{\Theta}{\delta \phi} \propto 2\sqrt{N_\textrm{p} N_\textrm{r}} \, \frac{N}{\zeta_\textrm{c}N + \zeta_\textrm{u}\sqrt{N} + 1} ,
\label{Eq:SNR}
\end{equation}
where $\zeta_\textrm{c}$, and $\zeta_\textrm{u}$, parameterize the strength of the correlated, and uncorrelated, classical noise, respectively, in units of the optical quantum noise. The size of these terms will be strongly dependent on the experimental realization and in practice both terms will be present to some extent. Examples of how the signal-to-noise ratio can scale is provided in Fig.~\ref{Fig-SNR}.

\subsection{Optical loss}

Using an $N$-cycle scheme has the additional advantage that one can now control the pulses within a sequence. Two example types of control include displacing the pulse to modify is magnitude and phase~\cite{Khosla2013} and changing the timing of the pulses to interact with a different mechanical quadrature. This can be used to better close the loop made in mechanical phase space in the presence of optical loss and thus reduce any unwanted residual correlations between the light and mechanics, which reduces the signal-to-noise ratio. In the following we describe a protocol to minimize the effects of optical loss (and related deleterious effects such as pulse shape distortion) without full control of the timing of the pulses as that would require an efficient, phase coherent, tunable optical delay line, which would introduce unwanted experimental complexity. In this model we neglect the added optical noise due to the effective optical self-Kerr effect, and the small change in the optical noise due to optical loss, as these effects will be small compared to the classical phase noise discussed above and the effects loss has on the mean of the optical field discussed below. Then, optical loss can be described by a change of the interaction parameter $\lambda$ after each pulsed interaction \cite{Pikovski2012_1}. For a single cycle, the four-pulse operator for regular quantum mechanics becomes
\begin{equation} \label{eqn:four-displacement_operator_noise}
\begin{split}
{\xi}_{\epsilon} & = e^{i\lambda_4 {n}_\textsc{l} {P}_\textsc{m}} e^{-i\lambda_3 {n}_\textsc{l} {X}_\textsc{m}} e^{-i \lambda_2 {n}_\textsc{l} {P}_\textsc{m}} e^{i \lambda_1 {n}_\textsc{l} {X}_\textsc{m}}  \\
& = e^{(\epsilon b^{\dagger} - \epsilon^* b)n_\textsc{l}} e^{-i \lambda_{\epsilon}^2 n_\textsc{l}^2} ,
\end{split}
\end{equation}
where $b^\dagger$ and $b$ are the mechanical creation and annihilation operators, respectively, and where
\begin{equation}
\begin{split} \label{eqn:epsilon}
\epsilon & = \frac{1}{\sqrt{2}} \left( \lambda_2 - \lambda_4 + i (\lambda_1 - \lambda_3)\right); \\
\lambda_{\epsilon}^2 & = \lambda_2 \lambda_3 + \frac{1}{2} \left(\lambda_1 - \lambda_3 \right) \left(\lambda_2 - \lambda_4 \right) .
\end{split}
\end{equation}
A varying interaction strength thus not only changes the strength of the signal, but also introduces an unwanted correlation between the light and the mechanical state. For an initial thermal state $\rho_\textsc{m}^{(th)}$ of the mechanics with the mechanical thermal occupation number $\bar{n}$, the mean phase becomes
\begin{equation}\label{eqn:mean_noise}
\begin{split}
\langle a \rangle_{\epsilon} & = \textrm{Tr}[a \xi_{\epsilon} | \alpha \rangle \langle \alpha | \otimes \rho_\textsc{m}^{(th)}  {\xi}_{\epsilon}^{\dagger} ] \\
& = \textrm{Tr}[ e^{-(\epsilon b^{\dagger} - \epsilon^* b)n_\textsc{l}} a \, e^{(\epsilon b^{\dagger} - \epsilon^* b)n_\textsc{l}} e^{-i \lambda_{\epsilon}^2 n_\textsc{l}^2} | \alpha \rangle \langle \alpha | \\
& \quad \otimes \rho_\textsc{m}^{(th)} e^{i \lambda_{\epsilon}^2 n_\textsc{l}^2} ] \\
& = \textrm{Tr}[ e^{i \lambda_{\epsilon}^2 n_\textsc{l}^2} a \,  e^{-i \lambda_{\epsilon}^2 n_\textsc{l}^2} | \alpha \rangle \langle \alpha | \otimes \rho_\textsc{m}^{(th)} e^{\epsilon b^{\dagger} - \epsilon^* b} ]\\
& = \textrm{Tr}_{\textsc{l}}[ a \,  e^{-i \lambda_{\epsilon}^2 (2 n_\textsc{l}+1)} | \alpha \rangle \langle \alpha |] \cdot \textrm{Tr}_{\textsc{m}}[\rho_\textsc{m}^{(th)} e^{\epsilon b^{\dagger} - \epsilon^* b}] \\
& = \langle a \rangle_{\textsc{qm}, \epsilon } \, e^{-\frac{|\epsilon|^2}{2}\left(1 +2 \bar{n} \right)} ,
\end{split}
\end{equation}
where we used the cyclic property of the trace and where $\langle a \rangle_{\textsc{qm}, \epsilon} = \bar\alpha e^{-i\lambda_{\epsilon}^2 - N_\textrm{p}\left(1-e^{-i2\lambda_{\epsilon}^2}\right)}$.
For $|\epsilon| \not = 0$, it can thus be beneficial to cool the mechanical mode of interest. For  the case of optical loss by a constant fraction in between each pulsed interaction, characterized by the intensity efficiency $\eta$ (where $\eta = 1$ is no loss), the strengths of each of the four displacements are given by $\lambda_1 = \lambda, \lambda_2=\eta \lambda, \lambda_3 = \eta^2 \lambda, \lambda_4 = \eta^3 \lambda$, and thus $\epsilon = \lambda (1 - \eta^2)(\eta + i)/\sqrt{2}$.

\begin{figure}[t]
\centering
\includegraphics[width=0.99\columnwidth]{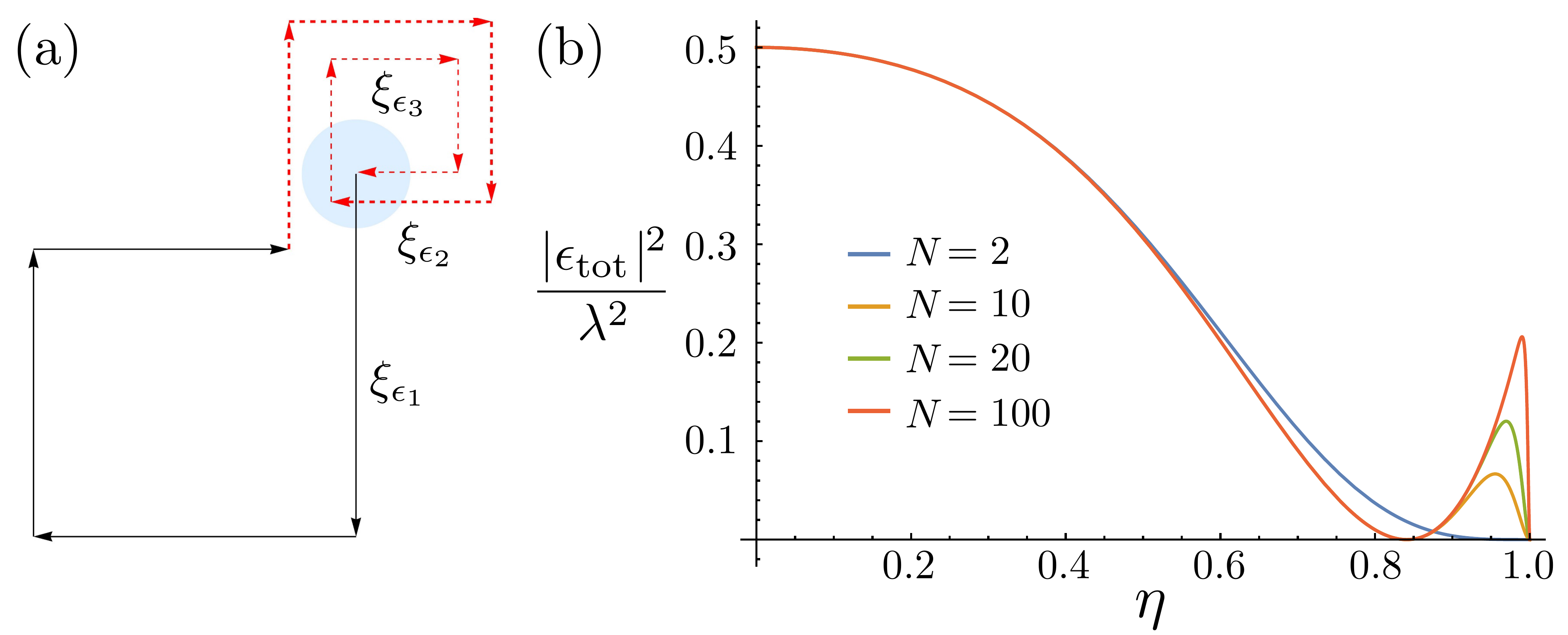}
\caption{\small (a) Displacement of the mechanical state (blue circle) using 3 cycles as described by Eq.~\eqref{Ndisplace}. Due to losses (here $\eta = 0.89$), the first loop (black arrows) does not close and thus causes degradation of the signal as given by Eq.~\eqref{eqn:mean_noise}. Two additional loops (red dashed arrows) bring the mechanical state back to its original position, thus giving $\epsilon_\textrm{tot}=0$ for this particular value of optical loss. (b) The parameter $|\epsilon_\textrm{tot}|^2$ as a function of optical loss $\eta$, as given in Eq.~\eqref{Ndisplace2}. A finite value of $|\epsilon_\textrm{tot}|^2$ reduces the signal.}
\label{Fig-Arrows}
\end{figure}

However, using a cycle of $N$ loops can reduce this deleterious effect, if one changes the direction of phase-space displacements after the first cycle. In the second  cycle (and all subsequent ones) one can apply the following modified four-displacement operator:
\begin{equation} \label{eqn:four-displacement_operator_noise2}
\begin{split}
{\xi}_{\epsilon_2} & = e^{-i\lambda_4^{(2)} {n}_\textsc{l} {P}_\textsc{m}} e^{i\lambda_3^{(2)} {n}_\textsc{l} {X}_\textsc{m}} e^{i \lambda_2^{(2)} {n}_\textsc{l} {P}_\textsc{m}} e^{-i \lambda_1^{(2)} {n}_\textsc{l} {X}_\textsc{m}}  \\
& = e^{-(\epsilon_2 b^{\dagger} - \epsilon^*_2 b) n_\textsc{l}} e^{-i \lambda_{\epsilon_2}^{2} n_\textsc{l}^2} .
\end{split}
\end{equation}
The sign of the correlating term is reversed, while the main phase has the same sign. This remains true for any commutator deformation that is of even power in the mechanical momentum. Thus, if the above sequence is performed after the sequence given by Eq. \eqref{eqn:four-displacement_operator_noise}, the second displacement is subtracted from the first displacement while the signal adds up as in the ideal case:
\begin{equation}
{\xi}_{\epsilon_2} {\xi}_{\epsilon_1} =  e^{((\epsilon_1-\epsilon_2) b^{\dagger} - (\epsilon_1-\epsilon_2)^* b)n_\textsc{l}} e^{-i (\lambda_{\epsilon_1}^{ 2} + \lambda_{\epsilon_2}^{ 2} ) n_\textsc{l}^2} .
\end{equation}
Here we assumed loss by a constant fraction as above, such that $\epsilon_2 = \eta^4 \epsilon_1$. In this case the additional phase factor vanishes, which is otherwise $e^{i 2 n_\textsc{l} \Im[\epsilon_1 \epsilon_2^*]}$.

We can repeat the same interaction multiple times. After $N$ cycles in total one has $\epsilon_N = \eta^{4(N-1)} \epsilon$ and $\lambda_{\epsilon_N}^{ 2} = \eta^{8(N-1)} \lambda_{\epsilon}^{2}$, where $\epsilon$ and $\lambda^2_\epsilon$ are given by Eq. \eqref{eqn:epsilon}. Thus, for $N-1$ further cycles with the sequence given in Eq.~\eqref{eqn:four-displacement_operator_noise2}, one obtains (without the commutator deformations):
\begin{equation}
\begin{split}
\xi_\textrm{tot} & = {\xi}_{\epsilon_N} ... {\xi}_{\epsilon_1} = e^{(\epsilon_\textrm{tot} b^{\dagger} - \epsilon_\textrm{tot}^* b)n_\textsc{l}}   e^{-i \lambda_\textrm{tot}^{ 2}  n_\textsc{l}^2} ,
\end{split} \label{Ndisplace}
\end{equation}
where
\begin{equation}
\begin{split}
\epsilon_\textrm{tot} & =   \epsilon \frac{1-2\eta^4+\eta^{4N}}{1-\eta^4} = \lambda (\eta + i) \frac{1-2\eta^4+\eta^{4N}}{\sqrt{2}(1+\eta^2)} ,  \\
\lambda_\textrm{tot} & = \lambda_{\epsilon}^{ 2} \sum_{j=0}^{N-1} \eta^{8j} = \eta \lambda^2 \frac{1 - \eta^{8 N}}{2(1-\eta^4)} .
\end{split}
\label{Ndisplace2}
\end{equation}
If $\eta \geq 2^{-1/4} \approx 0.84$, the unwanted correlating term can therefore be made to completely vanish, see Figure \ref{Fig-Arrows}. Depending on the value of $\eta$, the additional loops given by Eq.~\eqref{eqn:four-displacement_operator_noise2} may overcompensate for the losses, in which case some of the loops could be performed in the original direction as in Eq.~\eqref{eqn:four-displacement_operator_noise}. For $\eta < 0.84$, the effect of loss can still be reduced with the above scheme. In this case the optical phase is still affected as in Eq. \eqref{eqn:mean_noise}, but with a new effective parameter $|\epsilon|^2 \rightarrow |\epsilon_\textrm{tot}|^2$.\\

\subsection{Mechanical decoherence}

The mechanical resonator undergoes dissipation and decoherence during the protocol, as the mechanical mode is coupled to a thermal bath. This degrades the overall signal, which was estimated in Ref.~\cite{Pikovski2012_1} for a single cycle. For $N$ cycles, the mechanical decoherence will be increased as the system spends a longer time coupled to the bath between initialization of the protocol and final readout. Building on this earlier work, one can estimate the increase in decoherence to be proportional to $N$, as the overall coupling time to the bath is $2 \pi N/\omega_\textsc{m}$. This reduces the mean of the optical field to approximately $ \langle a \rangle \rightarrow \langle a \rangle \left[ 1 - \lambda^2 N k_\textrm{B} T/(\hbar \omega_\textsc{m} Q) \right]$, where $Q$ is the mechanical quality factor and $T$ is the temperature. This unwanted effect can be minimized by performing the experiment at cryogenic temperatures and using mechanical resonators with a high $\omega_\textsc{m} Q$ product. In contrast to other proposals that aim to directly study mechanical non-classicality, this protocol is more robust to mechanical thermal occupation and we do not anticipate that this effect will be a primary source of degradation to the signal-to-noise ratio.

\section{Discussion}

The enhancement to the signal-to-noise ratio provided by this scheme  can also help to constrain the applicability of the generalized uncertainty principle. As was noted in Ref.~\cite{Pikovski2012_1} and \cite{AmelinoCamelia2013_1}, it is an interesting open question which degrees of freedom are affected by the GUP and the expected signal will differ depending on where the model is applied. If the GUP commutators are applied to the `fundamental degrees of freedom', e.g. atoms or elementary particles, then the resultant GUP effects are reduced in comparison to application of the GUP deformations to the centre-of-mass motion. For example, applying the quadratic GUP to a composite system with $M$ fundamental constituents will result in a reduction in the signal by $M^k$, where $k$ lies between $1$ and $2$ depending on the correlations of the individual particles \cite{Pikovski2012_1}. With an $N$-cycle experimental scheme the GUP phase changes to $\Theta  \simeq \frac{4N}{3M^k} \bar{\gamma}^2 N_\textrm{p}^3 \lambda^4 e^{-i6\lambda^2}$, and thus, this reduction can be partially compensated by using multiple cycles. This discussion highlights that any experimental improvements provided by this scheme will not only help to determine the magnitude of any GUP correction to standard quantum mechanics but can also help shed light on the open question of what level such deformations may apply.



\section{Conclusions}

\label{concl}

Pulsed quantum optomechanics provides an exciting avenue to probe the generalized uncertainty principle (GUP) owing to its versatility to control of the motion of macroscopic mechanical oscillators and the high precision of optical interferometry. Here, we introduce a scheme to increase the sensitivity of measuring possible GUP deformations by using an extended pulse sequence of optomechanical interactions. Such an extended sequence allows a larger signal to be accumulated even for weak optomechanical coupling. We have identified and discussed several sources of technical noise including optical phase noise and optical loss as major contributions, and discussed how the advantage provided by this scheme diminishes with increasing amounts of such noise. Additionally, we have presented a strategy to dramatically reduce the effects of optical loss for a quadratic GUP model. These results help to shape the path ahead to experimentally explore some phenomenological models of quantum gravity on a table top.




\vspace{0.4cm}
\noindent
{\bf Acknowledgements}\\

P.B. and S.D. thank L. Spencer for discussions and E. C. Vagenas for pointing out some useful references. This work was supported in part by the Natural Sciences and Engineering Research Council of Canada; the NSF through a grant to ITAMP; the Society in Science Branco Weiss Fellowship, administered by the ETH Z\"{u}rich; and the Engineering and Physical Sciences Research Council (EP/N014995/1). M. R. V. would like to thank ITAMP for their kind hospitality provided during this project.\\

\providecommand{\abntreprintinfo}[1]{%
 \citeonline{#1}}

\begin{thebibliography}{}
\providecommand{\abntrefinfo}[3]{}
\providecommand{\abntbstabout}[1]{}
\abntbstabout{v-1.9.2 }


\bibitem{ACamelia1} G. Amelino-Camelia, {Living Rev. Rel.} {\bf 16}, 5 (2013).

\bibitem{Hossenfelder1} S. Hossenfelder, {Living Rev. Rel.} {\bf 16}, 2 (2013).

\bibitem{Plato2016} A. D. K. Plato, C. N. Hughes, and M. S. Kim, {Contemp. Phys.} {\bf 57}, 477 (2016).

\bibitem{expQGR}
G. Amelino-Camelia, {Nature} {\bf 398}, 216--218 (1999); D. Sudarsky, L. Urrutia and H. Vucetich. {Phys. Rev. Lett.} {\bf 89}, 231301 (2002); A. A. Abdo, et al.  Nature {\bf 462}, 331--334 (2009); A. S. Chou, et al. arXiv preprint (2015) [arXiv:1512.01216].


\bibitem{dv} S. Das, E. C. Vagenas, {Phys. Rev. Lett.} {\bf 101}, 221301 (2008).


\bibitem{Marin2013_1} F. Marin, et al. Nature Physics {\bf 9}, 71 (2013).

\bibitem{Marin2014} F. Marin, et al. New J. Phys.  {\bf 16}, 085012 (2014).

\bibitem{Bawaj2015} M. Bawaj, et al. Nature Communications {\bf 6}, 7503 (2015).

\bibitem{Bekenstein2012_1} J. D. Bekenstein, {Phys. Rev. D} {\bf 86}, 124040 (2012); Found. Phys. {\bf 44}, 452 (2014).


\bibitem{Garay}  L.~J.~Garay, {Int.\ J.\ Mod.\ Phys. A} {\bf 10}, 145 (1995).


\bibitem{Scardigli}
F.~Scardigli, {Phys.\ Lett. B} {\bf 452}, 39 (1999);


\bibitem{guppapers1} D. Amati, M. Ciafaloni, and G. Veneziano,
{Phys. Lett. B} {\bf 216}, 41 (1989).

\bibitem{cg} J. L. Cortes, J. Gamboa, {Phys. Rev. D} {\bf 71}, 065015 (2005).


\bibitem{maggiore}
M.~Maggiore, {Phys.\ Lett. B} {\bf 304}, 65 (1993);
  {Phys.\ Rev. D} {\bf 49}, 5182 (1994);
  {Phys.\ Lett. B} {\bf 319}, 83 (1993).

\bibitem{kmm} A. Kempf, {J. Math. Phys. } {\bf 35}, 4483 (1994);
    A. Kempf, G. Mangano, and R. B. Mann, {Phys. Rev. D} {\bf 52}, 1108 (1995);
    A. Kempf, and G. Mangano {Phys. Rev. D} {\bf 55}, 7909 (1997).

\bibitem{guppapers2}
S. Doplicher, K. Fredenhagen, and J. E. Roberts, {Commun. Math. Phys.} {\bf 172} 187-220 (1995);
S.~Hossenfelder, et al.,
  {Phys.\ Lett. B} {\bf 575}, 85 (2003);
C.~Bambi, and F.~R.~Urban, {Class.\ Quant.\ Grav. } {\bf 25}, 095006 (2008).

\bibitem{Ali2011_1}
A.~F. Ali, S.~Das, and E.~C. Vagenas, {Phys. Rev. D} {\bf 84}, 044013 (2011).

\bibitem{barun} B. Majumder, Phys. Lett. B {\bf 709}, 133 (2012).

\bibitem{ADV} A.~F.~Ali, S.~Das, and E.~C.~Vagenas, {Phys.\ Lett. B} {\bf 678}, 497 (2009).


\bibitem{Das2010_1} S.~Das, E.~C. Vagenas, and A.~F. Ali, {Phys. Lett. B} {\bf 690}, 407 (2010).

\bibitem{Deb2016_1}
S.~Deb, S.~Das, and E.~C. Vagenas, {Phys. Lett. B} {\bf 755}, 17 (2016).

\bibitem{Pikovski2012_1} I. Pikovski et al., {Nature Physics} {\bf 8}, 393 (2012).

\bibitem{Albrecht2014} A. Albrecht, A. Retzker, and M. B. Plenio, Phys. Rev. A \textbf{90}, 033834 (2014).

\bibitem{Rossi2016} M. A. C. Rossi, T. Giani, and M. G. A. Paris, Phys. Rev. D \textbf{94}, 024014 (2016).

\bibitem{Maggiore1993_2} M. Maggiore, {Phys. Lett. B}, {\bf 319}, 83 (1993).

\bibitem{Law1995_1} C. K. Law, Phys. Rev. A {\bf 51}, 2537 (1995).

\bibitem{Vanner2011} M. R. Vanner, et al. {Proc. Natl. Acad. Sci. USA} {\bf 108}, 16182-16187 (2011); M. R. Vanner, J. Hofer, G. D. Cole, M. Aspelmeyer, {Nature Communications} {\bf 4}, 2295 (2013).

\bibitem{Armata2016} F. Armata, et al. {Phys. Rev. A} {\bf 93}, 063862 (2016).

\bibitem{AmelinoCamelia2013_1}
G. Amelino-Camelia, {Phys. Rev. Lett.} {\bf 111}, 101301 (2013).


\bibitem{Khosla2013} K. E. Khosla, et al. {New J. Phys.} {\bf 15}, 043025 (2013).

\bibitem{Latmiral2016} L. Latmiral, F. Armata, M. G. Genoni, I. Pikovski, and M. S. Kim. {Phys. Rev. A} {\bf 93}, 052306 (2016).

\bibitem{BachorBook} H. A. Bachor and T. C. Ralph, \emph{A guide to experiments in quantum optics}, Wiley-VCH (2004).

\bibitem{VannerAnnalen} M. R. Vanner, I. Pikovski, and M. S. Kim, {Annalen der Physik} {\bf 527}, 15 (2015).

\end{thebibliography}

\end{document}